\documentclass[epj,nopacs]{svjour}

\usepackage{graphics}

\begin{document}
\def\makeheadbox{{%
    \hbox to0pt{\vbox{\baselineskip=10dd\hrule\hbox
        to\hsize{\vrule\kern3pt\vbox{\kern3pt
            \hbox{\emph{accepted for publication in Eur. Phys. J. A}}
            \kern3pt}\hfil\kern3pt\vrule}\hrule}%
      \hss}}}
\title{The {$^{198}$Au} $\beta^{-}$-half-life in the metal Au revisited}
\author{
  K. Fortak\inst{1}
  \and
  R. Kunz\inst{1}
  \and
  L. Gialanella\inst{2}
  \and
  H.-W. Becker\inst{3}
  \and
  J. Meijer\inst{1}
  \and
  F. Strieder\inst{4}
}
\mail{R. Kunz, \email{kunz@rubion.rub.de}}
\institute{
  RUBION, Ruhr-Universit\"at Bochum, Bochum, Germany
  \and
  INFN Sezione di Napoli, Napoli, Italy
  \and
  Fakult\"at f\"ur Physik und Astronomie, Ruhr-Universit\"at Bochum, Bochum, Germany
  \and
  Institut f\"ur Experimentalphysik, Ruhr-Universit\"at Bochum, Bochum, Germany
}
\date{Received: date / Revised version: date}
%
\abstract{
  The half-life of the $\beta^{-}$-decay of {$^{198}$Au} has been measured
  for room temperature and 12\,K. The resulting values of
  $T_{1/2}^{\rm (RT)} = 2.684 \pm 0.004\rm\,d$ and
  $T_{1/2}^{\rm (12\,K)} = 2.687 \pm 0.005\rm\,d$
  agree well within statistical uncertainties.
  An evidence for a temperature dependence of the half-life was not
  observed.
} 
\maketitle
In a recent experiment an increase in the half-life $T_{1/2}$
for the $\beta^{-}$\/-decay of {$^{198}$Au}
from $2.706\pm0.019\rm\,d$ to $2.802\pm0.020\rm\,d$
has been observed cooling the sample to $T=12\rm\,K$
\cite{Spillane:EPJA31:198Au}.
This observation triggered a number of half-life measurements
on various isotopes, which could not confirm the results
\cite{Kumar:PRC77:198Au7Be,Goodwin:EPJA34:198Au,
  Ruprecht:PRC77:22Na198Au196Au,Ruprecht:PRC78:22Na198Au196Au:err,
  Farkas:JPhysG36:74As,Goodwin:97Ru}.

Further measurements have been undertaken to resolve
this discrepancy, using a similar setup as described
by Spillane \emph{et al.} \cite{Spillane:EPJA31:198Au}

The experimental setup consisted of a vacuum chamber equipped with a
turbomolecular pump and two cryo\-pumps. The gold sample was installed
on top of the cooling rod of one of the cryo\-pumps. The pressure
reached within the chamber was lower than $10^{-9}$\,mbar.  The
temperature at the sample was measured using a Silicon diode (Lake
Shore Cryotronics DT-670).

A high-purity germanium detector (relative efficiency 120\% at
$E_{\gamma} = 1.33$\,MeV) was placed at $0^{\circ}$ with respect to the
cryopump-axis as in Spillane \emph{et al.} \cite{Spillane:EPJA31:198Au}
The distance between the detector and the Au-Foil was about 2\,cm
to improve statistics compared to the previous experiment.

In the present experiment the same Au-foils as in
\cite{Spillane:EPJA31:198Au} were used (thickness
$=0.5$\,mm, area $=2\times 2$\,cm$^2$,
impurities less than
1\,ppm O and H; obtained from Chempur). They were activated
via the reaction $^{197}\mbox{Au}\,(\mbox{n},\gamma)^{198}\mbox{Au}$
using the AmBe neutron source at the Isotopenlabor of
Ruhr-Universit\"at Bochum.
Saturation is achieved approximately within one week.
In total three runs have been performed at room temperature,
and at $T=12\rm\,K$, respectively.

To measure the decay curve the 412\,keV $\gamma$-rays were observed,
that are emitted by {$^{198}$Hg} after the $\beta^{-}$-decay of {$^{198}$Au}.
The runtime for each spectrum was one hour.

\begin{figure*}
  \resizebox{0.49\linewidth}{!}{%
    \includegraphics{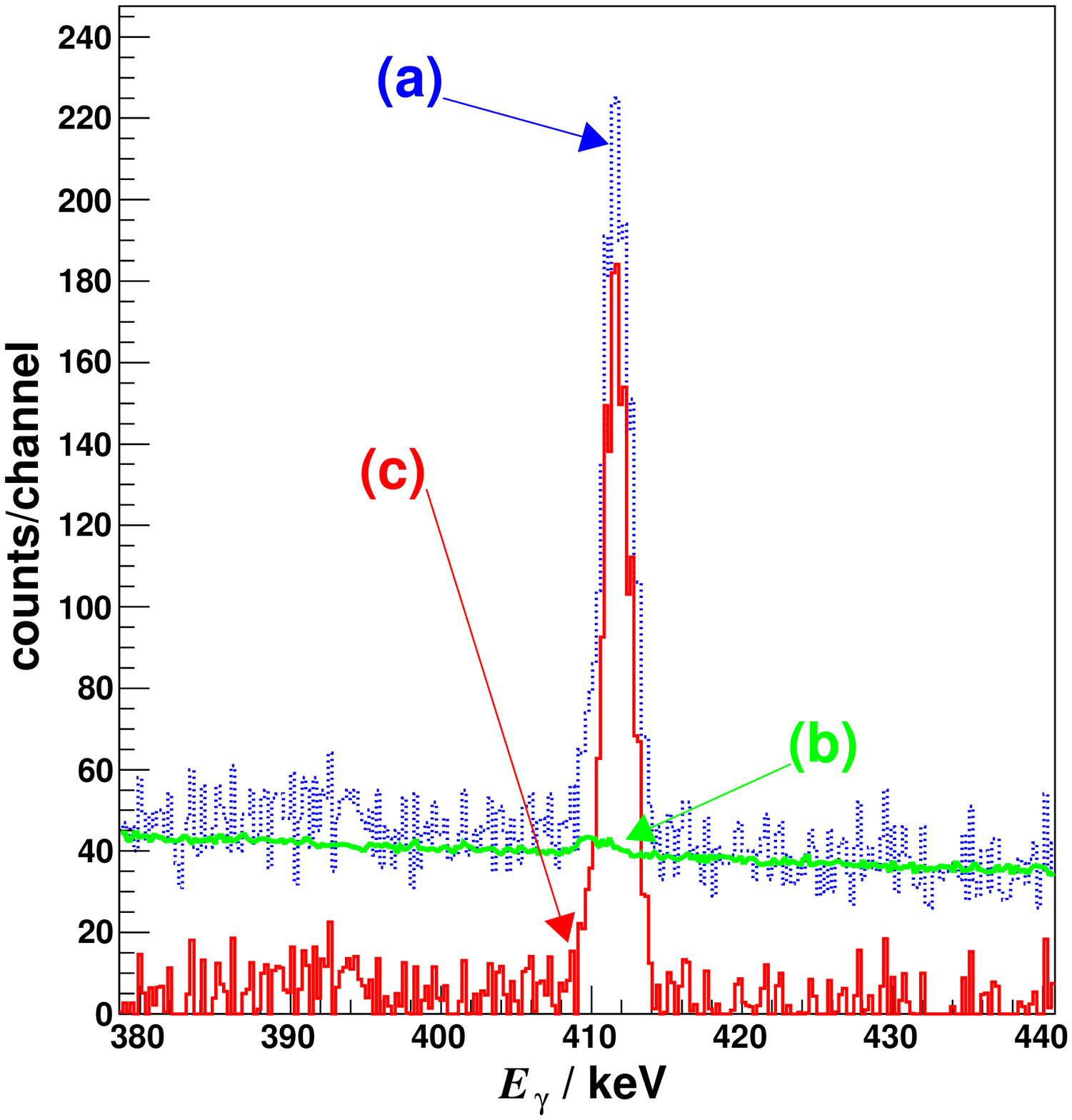}
  }%
  \hfill%
  \resizebox{0.49\linewidth}{!}{%
    \includegraphics{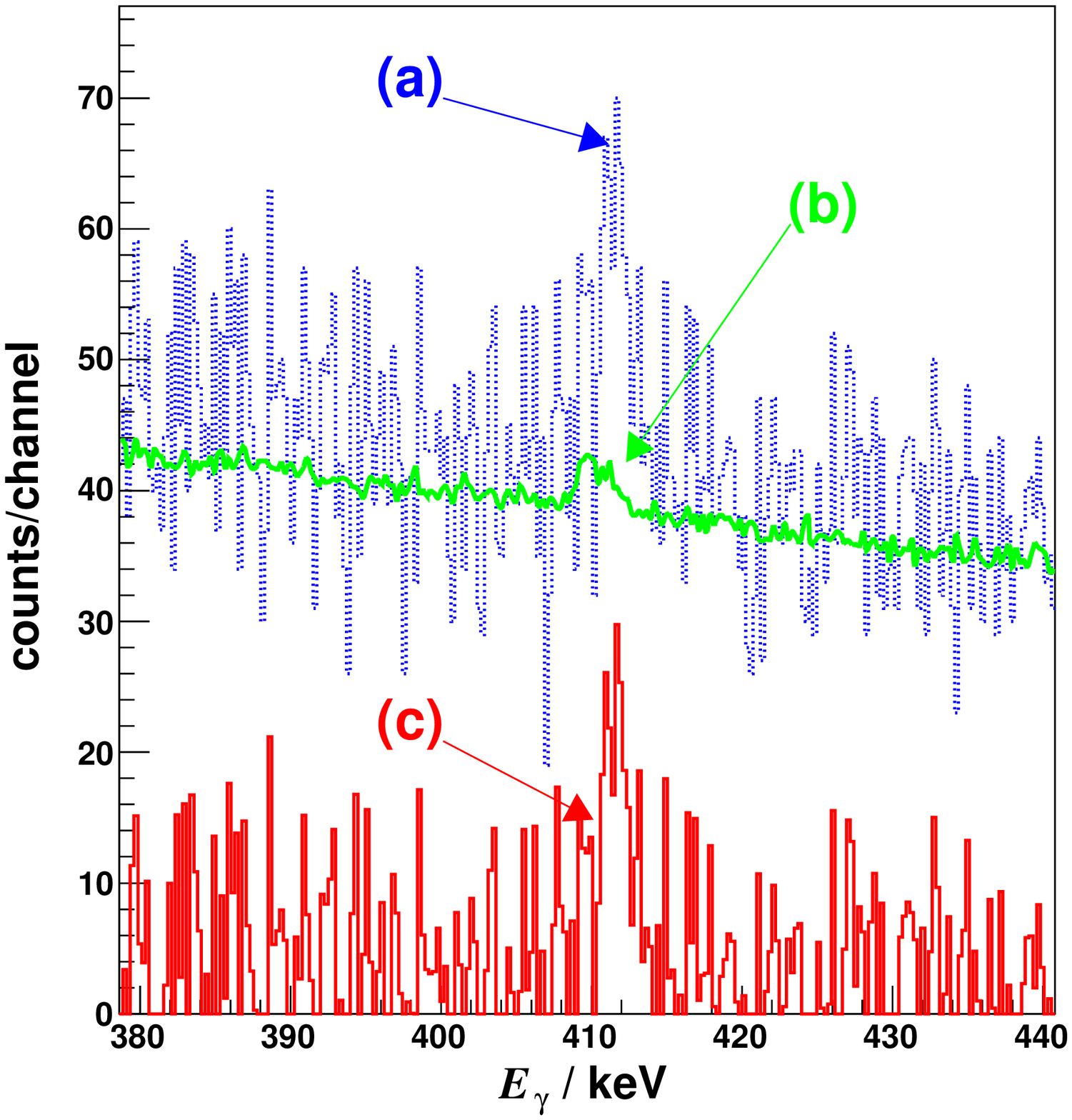}
  }
  \caption{Examples for the subtraction of the background spectrum in the region
    of the 412\,keV $\gamma$-line. In the left panel the spectrum for $t=100\rm\,h$
    is shown, in the right panel for $t=300\rm\,h$. In each graph curve (a) is the original
    spectrum (runtime 1 hour), curve (b) is the background spectrum with a runtime
    of 143 hours, normalized to the runtime of the original spectrum using the
    counting rate in the $E_{\gamma}=1461\rm\,keV$ peak of {$^{40}$K}.
    Curve (c) shows the resulting spectrum after subtraction of the normalized
    background spectrum.
    The small peak in the background spectrum at 409.5\,keV results from
    a $\gamma$-transition in {$^{228}$Th} following the $\beta^{-}$-decay of {$^{228}$Ac},
    belonging to the natural Thorium series.
  }
  \label{fig:spec1}
\end{figure*}

Additional background spectra with one hour runtime each
were taken over several days before and after the measurements with the
sample.  The resulting background spectrum has been determined by
summing up all background spectra from one series. This sum spectrum
has been subtracted from each Au-spectrum after normalization.
Two examples for this procedure are
shown in fig.~\ref{fig:spec1} for the relevant region around
$E_{\gamma}=412\rm\,keV$. In the background, a small peak at 409.5\,keV
is visible due to a $\gamma$-transition in {$^{228}$Th} following the
$\beta^{-}$-decay of {$^{228}$Ac}, belonging to the natural Thorium
series.
This $\gamma$-ray line can only be clearly identified in very high
statistics background runs, i.e. here a 142 hour run.
However, this line may have an effect on the {$^{198}$Au} $\gamma$-ray line
analysis at low counting rate and, hence, on the extracted half-life.

After mounting the Au-sample at the cooling rod of the cryopump, the
turbopump was started first to reach a vacuum of less than
$10^{-7}$\,mbar. During the room temperature measurements, only the
turbopump was used.
For the measurements at 12\,K,
both cryopumps were started consecutively,
to avoid any contamination of the sample by deposition of rest gas atoms.

The {$^{40}$K} line at $E_{\gamma}=1461\rm\,keV$ has been used
to normalize all spectra and to correct for deadtime effects.
An additional $50\rm\,Hz$ pulser has been used in selected runs to
obtain the counting rate in the {$^{40}$K} line.

The peak area of the 412\,keV $\gamma$-ray line has been determined
by integration of the natural background subtracted spectrum.
The remaining background around the peak has been interpolated
with a linear function and has been subtracted.
One should note that without prior
background subtraction the evaluation of the half-life has
proven to be very sensitive to the choice of the background
area around the peak.

The resulting activities have been fitted using the usual
decay formula
\begin{equation}
  \label{eq:decay}
  A(t) = A_{0}\,\exp\left(-\lambda\,t\right)
\end{equation}
where
\begin{equation}
  \label{eq:lambda}
  \lambda=\frac{\ln2}{T_{1/2}}
\end{equation}
is the decay constant and $A_{0}$ is the initial activity for $t=0$.

Due to the relatively short half-life, normally
eq.~\ref{eq:decay} has to be integrated over the
time $\Delta{}t$ of a single run starting at $t_{0}$.
In the present experiment $\Delta{}t$ varies only due to negligible
differences in the dead time and eq.~\ref{eq:decay} can be used
except for a modified initial activity.
The half-life is independent of the absolute activity,
hence, the final result is not influenced by this modification.

\begin{table*}
  \caption{List of all measurements and the resulting fit parameters with their
    statistical uncertainties and the reduced chi-square.
    RT denotes room temperature, $\Delta{}T$ is the total run time of the measurement.
    In the right part the decay constant was fixed to the literature value
    $\lambda_{\rm lit.} = \frac{\ln2}{T_{1/2, \rm lit.}} = 0.25714\rm\,d^{-1}$,
    which also gives excellent fits.
  }
  \label{tab:fitcmp}
  \begin{tabular}{ccc|r@{$\,\pm\,$}lr@{$\,\pm\,$}lr@{$\,\pm\,$}lc|r@{$\,\pm\,$}lc}
    \hline
    &
    &
    & \multicolumn{7}{c|}{fit with $\lambda$ free}
    & \multicolumn{3}{c}{fit with $\lambda=\lambda_{\rm lit.}$ fixed}
    \\
    \cline{4-13}
    Run & $T$ & $\Delta T$ (h)
    & \multicolumn{2}{c}{$A_{0}\rm~(s^{-1})$}
    & \multicolumn{2}{c}{$\lambda\rm~(d^{-1})$}
    & \multicolumn{2}{c}{$T_{1/2}\rm~(d)$}
    & $\chi^{2}_{\rm red.}$
    & \multicolumn{2}{c}{$A_{0}\rm~(s^{-1})$}
    & $\chi^{2}_{\rm red.}$\vphantom{\raisebox{3pt}{M}}
    \\
    \hline
    1 & RT & 266
    & 11.14 & 0.04 & 0.2581 & 0.0007 & 2.686 & 0.007
    & 0.9400
    & 11.087 & 0.021
    & 0.9456
    \\
    2 & RT & 119
    & 5.180 & 0.022 & 0.2576 & 0.0016 & 2.691 & 0.016
    & 1.0351
    & 5.174 & 0.012
    & 1.0273
    \\
    3 & 12 K & 286
    & 1.070 & 0.006 & 0.2559 & 0.0016 & 2.709 & 0.017
    & 0.9181
    & 1.074 & 0.004
    & 0.9172
    \\
    4 & 12 K & 307
    & 1.005 & 0.006 & 0.2576 & 0.0017 & 2.691 & 0.017
    & 0.9494
    & 1.004 & 0.004
    & 0.9465
    \\
    5 & RT & 313
    & 8.235 & 0.023 & 0.2584 & 0.0005 & 2.683 & 0.006
    & 0.9692
    & 8.187 & 0.013
    & 0.9874
    \\
    6 & 12 K & 502
    & 6.376 & 0.018 & 0.2581 & 0.0005 & 2.685 & 0.006
    & 0.9171
    & 6.347 & 0.011
    & 0.9239
    \\
    \hline
  \end{tabular}
\end{table*}

\begin{figure}
  \resizebox{\linewidth}{!}{%
    \includegraphics{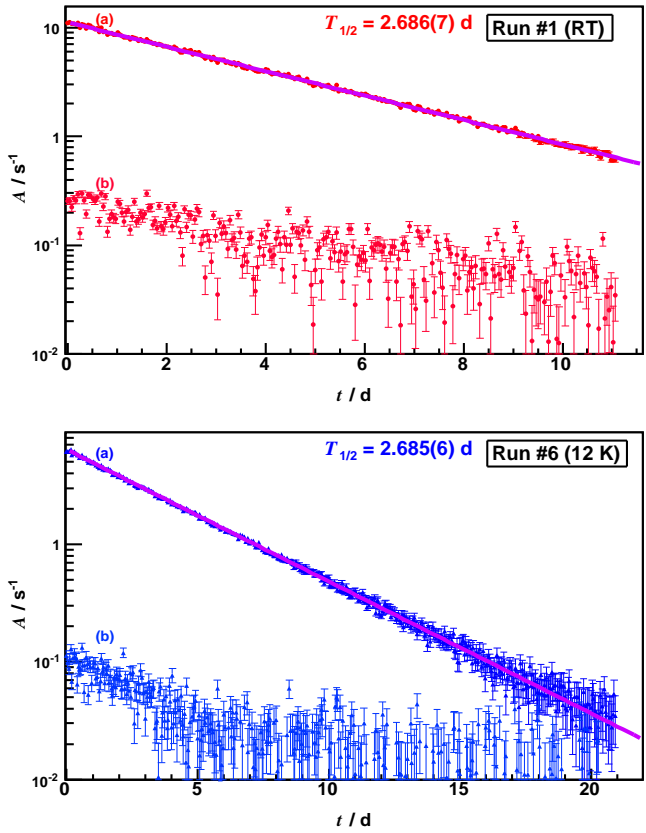}
  }
  \caption{Decay curves at room temperature (``RT'')
    and for $T=12\rm\,K$.
    (a) shows the decay curve where the peak area is evaluated by integrating
    the region of the gamma-peak after subtracting in a first step
    the corresponding environmental background spectrum and
    in a second step the remaining background determined in areas left and
    right of the peak (width normalized to the width of the peak region).
    The data points (b) show this remaining normalized background.
  }
  \label{fig:decay}
\end{figure}

Finally, each decay curve has been fitted using the
initial activity $A_{0}$ and the decay constant $\lambda$ as free
parameters. Additional fits have been performed fixing
$\lambda$ at the literature value $\lambda_{\rm lit.}$, only varying $A_{0}$,
to test the consistency of the experiment with the literature
value of $\lambda$.
The resulting parameters with their statistical uncertainties
are shown in table~\ref{tab:fitcmp} along
with the reduced chi-square $\chi^{2}_{\rm red.}$.
Fig.~\ref{fig:decay} shows decay curves
for room temperature and 12\,K respectively.

\begin{figure}
  \resizebox{\linewidth}{!}{%
    \includegraphics{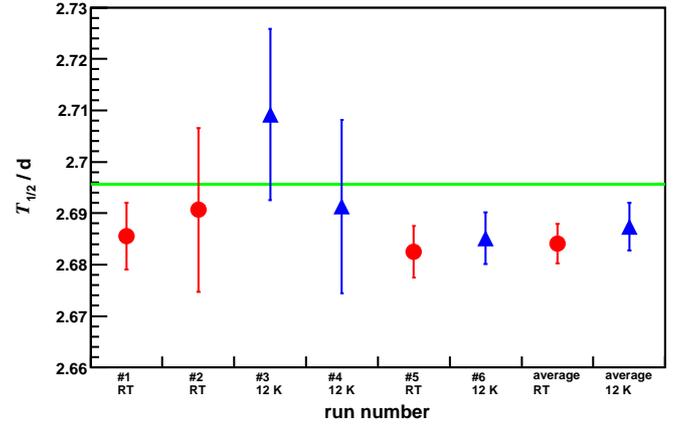}
  }
  \caption{Comparison of all obtained half-lifes in the order of measurements.
    Experiments at room temperature are represented by dots,
    experiments at $T=12\rm\,K$ with triangles.
    The two rightmost points represent the average values.
    The error bars only include statistical uncertainties.
    The horizontal bar indicates the literature value,
    $T_{1/2,\rm lit.} = 2.6956\pm0.0003\rm\,d$, with its uncertainty.
  }
  \label{fig:T12overview}
\end{figure}

As shown in table~\ref{tab:fitcmp}, all fits are compatible
when the decay constant has been fixed at the literature value;
using it as a free parameter gives only a slight improvement in
$\chi^2_{\rm red}$, if any.
An overview of the half-life results is shown in
table~\ref{tab:halflife} and fig.~\ref{fig:T12overview}
together with the mean values of $T_{1/2}$
for room temperature and $12\rm\,K$.
The latter values have been calculated including
statistical errors only.
In conclusion, the averaged half-lifes
$T_{1/2}^{\rm (RT)} = 2.684 \pm 0.004\rm\,d$ at room temperature
and
$T_{1/2}^{\rm (12\,K)} = 2.687 \pm 0.005\rm\,d$ at 12\,K
show no effect of a change within the statistical uncertainties.

Thus, the present experiment excludes the large effect observed
previously \cite{Spillane:EPJA31:198Au}.
A possible explanation for the assumed change in half-lifes can
be given by the difficulty analyzing the spectra without
subtraction of a high statistic background spectrum.
In the present analysis the obtained half-life was affected strongly
by the choice of the region around the 412\,keV $\gamma$-ray line
used for the background determination.
This effect vanished when the background spectrum was subtracted
before the peak was evaluated.
Particularly a proper background measurement and subtraction
of the 409.5\,keV $\gamma$-ray line is important
before the peak is integrated.

In summary we confirm the conclusion of
recent works \cite{Kumar:PRC77:198Au7Be,Goodwin:EPJA34:198Au,
  Ruprecht:PRC77:22Na198Au196Au,Ruprecht:PRC78:22Na198Au196Au:err}
that no temperature effect on the half-life of
{$^{198}$Au} is present within the statistical uncertainties
of this experiment.

\begin{table}
  \caption{Half-lifes for all runs, ordered by temperature.
    The resulting average values for room temperature and
    $T=12\rm\,K$ are in good agreement.
    The values for the single runs are given with higher
    accuracy than required by the errors, because reducing
    this accuracy leads to a change in the average values
    in the last significant digit by 1.
    As usual, uncertainties have been rounded up.
  }
  \label{tab:halflife}
  \begin{tabular}{cc@{$\,\pm\,$}c}
    \hline\noalign{\smallskip}
    \multicolumn{3}{l}{Room temperature}
    \\
    \noalign{\smallskip}\hline\noalign{\smallskip}
    Run & \multicolumn{2}{c}{$T_{1/2}\rm\,(d)$} \\
    \noalign{\smallskip}\hline\noalign{\smallskip}
    1 & 2.6855 & 0.0065 \\
    2 & 2.6906 & 0.0160 \\
    5 & 2.6825 & 0.0051 \\
    \noalign{\smallskip}\hline\noalign{\smallskip}
    average & 2.684 & 0.004
    \\
    \noalign{\smallskip}\hline\noalign{\smallskip}
  \end{tabular}
  \quad
  \begin{tabular}{cc@{$\,\pm\,$}c}
    \hline\noalign{\smallskip}
    \multicolumn{3}{l}{$T=12\rm\,K$}
    \\
    \noalign{\smallskip}\hline\noalign{\smallskip}
    Run & \multicolumn{2}{c}{$T_{1/2}\rm\,(d)$} \\
    \noalign{\smallskip}\hline\noalign{\smallskip}
    3 & 2.7092 & 0.0167 \\
    4 & 2.6913 & 0.0169 \\
    6 & 2.6851 & 0.0051 \\
    \noalign{\smallskip}\hline\noalign{\smallskip}
    average & 2.687 & 0.005 \\
    \noalign{\smallskip}\hline\noalign{\smallskip}
  \end{tabular}
\end{table}


\begin{thebibliography}{}
\bibitem{Spillane:EPJA31:198Au}
  T.~Spillane \emph{et al.},
  Eur. Phys. J. A \textbf{31}, 203 (2007).
  doi: 10.1140/epja/i2006-10212-8.
\bibitem{Kumar:PRC77:198Au7Be}
  V.~Kumar \emph{et al.},
  Phys. Rev. C \textbf{77}, 051304(R) (2008).
  doi: 10.1103/PhysRevC.77.051304.
\bibitem{Goodwin:EPJA34:198Au}
  J.~R. Goodwin \emph{et al.},
  Eur. Phys. J. A \textbf{34}, 271 (2007).
  doi: 10.1140/epja/i2007-10509-0.
\bibitem{Ruprecht:PRC77:22Na198Au196Au}
  G.~Ruprecht \emph{et al.},
  Phys. Rev. C \textbf{77}, 065502 (2008).
  doi: 10.1103/PhysRevC.77.065502.
\bibitem{Ruprecht:PRC78:22Na198Au196Au:err}
  G.~Ruprecht \emph{et al.},
  Phys. Rev. C \textbf{78}, 039901(E) (2008).
  doi: 10.1103/PhysRevC.78.039901.
\bibitem{Farkas:JPhysG36:74As}
  J.~Farkas \emph{et al.},
  J. Phys. G: Nucl. Part. Phys. \textbf{36}, 105101 (2009).
  doi: 10.1088/0954-3899/36/10/105101.
\bibitem{Goodwin:97Ru}
  J.~R. Goodwin \emph{et al.},
  Phys. Rev. C \textbf{80}, 045501 (2009).
  doi: 10.1103/PhysRevC.80.045501.
\end{thebibliography}
\end{document}